\documentclass[leqno, 12pt]{article}
\usepackage{times}
\usepackage{amsmath}
\usepackage{amssymb}
\usepackage{setspace}

\numberwithin{equation}{section}

\newtheorem{theorem}{Theorem}[section]
\newtheorem{definition}[theorem]{Definition}

\newtheorem{proposition}[theorem]{Proposition}
\newtheorem{lemma}[theorem]{Lemma}
\newtheorem{corollary}[theorem]{Corollary}

\begin{document}

\date{}

\title{A modal logic amalgam of classical and intuitionistic logic}

\author{Steffen Lewitzka\thanks{Universidade Federal da Bahia -- UFBA,
Instituto de Matem\'atica,
Departamento de Ci\^encia da Computa\c c\~ao,
Campus de Ondina,
40170-110 Salvador -- BA,
Brazil,
e-mail: steffen@dcc.ufba.br}}

\maketitle

\begin{abstract}
A famous result, conjectured by G\"odel in 1932 and proved by McKinsey and Tarski in 1948, says that $\varphi$ is a theorem of intuitionistic propositional logic IPC iff its G\"odel-translation $\varphi'$ is a theorem of modal logic S4.  In this paper, we extend an intuitionistic version of modal logic S1+SP, introduced in our previous paper \cite{lewjlc}, to a classical modal logic L and prove the following: a propositional formula $\varphi$ is a theorem of IPC iff $\square\varphi$ is a theorem of L (actually, we show: $\Phi\vdash_{IPC}\varphi$ iff $\square\Phi\vdash_L\square\varphi$, for propositional $\Phi,\varphi$). Thus, the map $\varphi\mapsto\square\varphi$ is an embedding of IPC into L, i.e. L contains a copy of IPC. Moreover, L is a conservative extension of classical propositional logic CPC. In this sense, L is an amalgam of CPC and IPC. We show that L is sound and complete w.r.t. a class of special Heyting algebras with a (non-normal) modal operator.
\end{abstract}


\section{Introduction}

According to the informal Brouwer-Heyting-Kolmogorov semantics (BHK) of intuitionistic propositional logic (IPC), intuitionistic truth is provability: a formula is \textit{true} if there is a proof for it. Logical connectives then have a constructive meaning. For instance, a proof of $\varphi\vee\psi$ is given by a proof of $\varphi$ or by a proof of $\psi$. In an attempt to formalize BHK, G\"odel \cite{goe} interprets IPC in a modal extension of classical propositional logic (namely Lewis system S4) by defining a \textit{translation} $\varphi\mapsto\varphi'$ that maps any propositional formula $\varphi$ to a modal formula $\varphi'$ such that the following holds: if $\varphi$ is a theorem of IPC, then $\varphi'$ is a theorem of S4. G\"odel also conjectured the converse, i.e. 
\begin{equation}\label{Goedel}
\vdash_{IPC}\varphi\text{ }\Leftrightarrow\text{ }\vdash_{S4}\varphi'
\end{equation}
for any propositional $\varphi$. This conjecture was later proved by McKinsey and Tarski \cite{mcktar}.\footnote{For much more details and historical background, we refer the reader to \cite{ottfei, artbek}.} Thus, Brouwer's intuitionistic logic, as axiomatized by Heyting, can be recovered from modal system S4 by the equivalence \eqref{Goedel} above. However, since the G\"odel translation $\varphi\mapsto\varphi'$ is not trivial, S4 contains IPC only in indirect, codified form. In particular, the modal operator of system S=S4 cannot be seen as a predicate for intuitionistic truth in the following sense:
\begin{equation}\label{truth}
\square\varphi\text{ is true in S iff }\varphi\text{ belongs to a given prime theory of IPC}
\end{equation}
for any propositional $\varphi$.\footnote{Of course, \textit{truth} is a semantic notion which is defined relative to a given model. Recall that the set of formulas true at a given possible world of a Kripke model of intuitionistic logic forms a prime theory, and each prime theory corresponds to a world of a Kripke model. Hence, in intuitionistic logic, ``$\varphi$ is true" means that $\varphi$ belongs to a given prime theory of IPC.}

In this paper, we present a modal logic L such that \eqref{truth} above holds with S=L. Soundness of L then implies the right-to-left direction of the following equivalence \eqref{Lew}. The left-to-right direction can easily be shown by an induction on the length of a derivation. For any propositional $\varphi$: 
\begin{equation}\label{Lew}
\vdash_{IPC}\varphi\text{ }\Leftrightarrow\text{ }\vdash_{L}\square\varphi
\end{equation}
Actually, we will show the following stronger result:
\begin{equation}\label{Mirror}
\Phi\vdash_{IPC}\varphi\text{ }\Leftrightarrow\text{ }\square\Phi\vdash_{L}\square\varphi,
\end{equation}
for any set of propositional formulas $\Phi\cup\{\varphi\}$. Thus, derivations in IPC are mirrored by corresponding derivations in L by means of the modal operator. In particular, the mapping $\varphi\mapsto\square\varphi$ defines an embedding of IPC into L. That is, L contains a copy of IPC and behaves in a similar way as a conservative extension. Moreover, L is a conservative extension of classical propositional logic CPC. In this sense, L can be viewed as an amalgam or a combination of IPC and CPC. In the combined logic L, IPC is separated from CPC by means of the modal operator, which avoids the collapsing of both logics. Combinations of logics in a much more general context have been studied extensively over the last years (see \cite{carcon} for an excellent overview). Known combining techniques largely rely on the concept of \textit{fibring} originally introduced by Gabbay (see \cite{gab1, gab2, gab3}). These techniques generally assume that the object logics are defined over different languages. The so-called \textit{collapsing problem} was first identified in \cite{gab1, gab2} for the special case of combining intuitionistic and classical propositional logic: in the semantics of the combined logic, the logical connectives of the composed language have classical behaviour. That is, the combination results in a collapsing of intuitionistic logic into classical logic. A logical system that combines classical and intuitionistic logic avoiding the problem is found in \cite{cerher}. A first general solution to the collapsing problem, based on \textit{modulated fibring}, is presented in \cite{carserras}. In \cite{calram}, the authors propose \textit{cryptofibring semantics} as a generalization of fibring semantics and show that this provides a solution to the collapsing problem. Other solutions to the problem are given by \textit{graph-theoretic fibring} \cite{serserrascon1, serserrascon2} and, more recently, by a new method called \textit{meet combination of logics} \cite{serserras1, serserras2}. 

The present paper is not in the tradition of \textit{fibring} or related techniques of combining logics, neither it pretends to compete with the sophisticated techniques developed in that research line. In fact, our approach relies on different assumptions and motivations, which makes a direct comparison to complex fibring techniques a difficult task. One of the main differences is the fact that our object logics IPC and CPC are given in exactly the same propositional language. In the combined system, the object logics then are distinguished by a modal operator. Semantically, the intuitionistic part of the amalgam is, roughly speaking, given by special Heyting algebras whereas the classical part is given by ultrafilters contained in those Heyting algebras. This represents a very compact semantical solution which, together with the mentioned syntactical simplifications, ensures that, in our view, the approach is technically simpler and less complex than known fibring techniques. We believe that the proposed solution can be adapted to similar cases of combining logics. To what extent this can be done and which are the limitations in comparison to fibring techniques remains to be further investigated. We would like to point out that the original motivation for the present approach was not to develop new combining techniques but rather to find a classical modal logic L with the properties \eqref{truth} -- \eqref{Mirror} above. Of course, it would be nice to have a \textit{minimal} logic with those properties. Lewis modal system S1 turns out to be a promising base. In our previous paper \cite{lewjlc}, we found an algebraic semantics for the slightly stronger system S1+SP. The semantics essentially relies on principles of non-Fregean logic \cite{blosus} paired with the idea to identify strict equivalence $\square(\varphi\leftrightarrow\psi)$ with propositional identity $\varphi\equiv\psi$. That is, we define an identity connective by $\varphi\equiv\psi := \square(\varphi\leftrightarrow\psi)$. If a denotational semantics is available, then an identity connective should fulfill the following condition: $\varphi\equiv\psi$ is satisfied in a model iff $\varphi$ and $\psi$ denote the same proposition, i.e. the same element of the underlying model-theoretic universe. In this paper, we show that it is enough to extend an intuitionistic version of S1+SP by an axiom scheme that represents a certain \textit{disjunction property} and a scheme of theorems that represents the classical principle of \textit{tertium non datur} in order to obtain a logic L with the desired properties.

\section{Modal logic L}

The set $Fm$ of formulas is inductively defined in the usual way over an infinite set $V=\{x_0,x_1,x_2,...\}$ of propositional variables, logical connectives $\bot,\rightarrow,\vee,\wedge$, and the modal operator $\square$. If $x$ is a variable and $\varphi,\psi$ are formulas, then we write $\varphi[x:=\psi]$ for the formula that results from substituting $\psi$ for $x$ in $\varphi$. By $Fm_0$ we denote the set of (non-modal) propositional formulas, i.e. formulas without modal operator $\square$. We use the following abbreviations:
\begin{itemize}
\item $\varphi\leftrightarrow\psi :=(\varphi\rightarrow\psi)\wedge(\psi\rightarrow\varphi)$
\item $\neg\varphi :=\varphi\rightarrow\bot$
\item $\top :=\neg\bot$
\item $\varphi\equiv\psi :=\square(\varphi\rightarrow\psi)\wedge \square(\psi\rightarrow\varphi)$ (strict equivalence)
\item $\square\Phi:=\{\square\varphi\mid\varphi\in\Phi\}$, for any set $\Phi\subseteq Fm$.
\end{itemize}

In particular, we define an identity connective by strict equivalence. \\

In current logics, the meaning (denotation, \textit{Bedeutung}) of a formula remains unchanged if we replace a subformula by a formula with the same meaning. Actually, this property represents a general ontological law sometimes called in the literature the \textit{Indiscernibility of Identicals}. In a propositional language with (definable) connectives for identity and implication, that law can be expressed by the following Substitution Principle SP:
\begin{equation*}
(\varphi\equiv\psi)\rightarrow(\chi[x:=\varphi]\equiv\chi[x:=\psi])
\end{equation*}

If propositional identity $\varphi\equiv\psi$ is given by strict equivalence, then most of current modal logics (including non-normal modal logic S3, see \cite{lewjlc}) satisfy SP. An exception is Lewis modal system S1, where that law is fulfilled in a weaker form, namely as the rule of \textit{Substitution of Proved Strict Equivalents} (SPSE): ``If $\varphi\equiv\psi$ is a theorem, then any formula of the form $\chi[x:=\varphi]\equiv\chi[x:=\psi]$ is a theorem". There is no known intuitive semantics for S1, and the main reason for that fact seems to be the weakness of rule SPSE. Indeed, in \cite{lewjlc} we define a natural algebraic semantics for modal system S1+SP, i.e the logic that results from S1 by replacing rule SPSE with the stronger scheme of theorems SP.\\

The axioms are given by the following schemes (i)--(iv):
\begin{enumerate}
\item intuitionistic propositional tautologies and their substitution-instances\footnote{By a substitution-instance of a formula $\varphi\in Fm_0$, we mean a formula which results from $\varphi$ by replacing some of its variables by formulas.} 
\item $\square\varphi\rightarrow\varphi$
\item $\square(\varphi\rightarrow\psi)\rightarrow(\square(\psi\rightarrow\chi)\rightarrow\square(\varphi\rightarrow\chi))$
\item $\square(\varphi\vee\psi)\rightarrow(\square\varphi\vee\square\psi)$ (disjunction property)
\end{enumerate}

The inference rules are:
\begin{itemize}
\item Modus Ponens MP: ``From $\varphi$ and $\varphi\rightarrow\psi$ infer $\psi$."
\item Axiom Necessitation AN: ``If $\varphi$ is an axiom of scheme (i)--(iv), then infer $\square\varphi$."
\end{itemize}

Logic L is given by the above system of axiomes and rules \textit{plus} theorems of the form $\varphi\vee\neg\varphi$ and SP. We write $\Phi\vdash_L\varphi$ or shorter $\Phi\vdash\varphi$ if there is a derivation of $\varphi$ from $\Phi$ in logic L. (Rule AN applies only to axioms but not to theorems.)\\
Let (i)' be as (i) above, but with all classical tautologies instead of only intuitionistic ones. Then logic S1+SP, as introduced in \cite{lewjlc}, is given by the system of axioms (i)',(ii) and (iii), rules MP and AN, and all formulas of the form SP as theorems.\\  
Note that because of scheme (iv), L is contained in no Lewis modal system. In fact, it might be hard to find any natural Kripke-style semantics where (iv) would be valid.\\

The Deduction Theorem can be proved in the usual way by induction on the length of a derivation. Also the following results will be useful. Their proofs can be found in \cite{lewjlc} (Lemma 2.3 and Lemma 2.4, respectively).

\begin{lemma}\label{124}\cite{lewjlc}
For all formulas $\varphi, \psi$:
\begin{itemize}
\item $\vdash\square\varphi\leftrightarrow(\varphi\equiv\top)$
\item  $\vdash\square(\varphi\rightarrow\psi)\rightarrow(\square\varphi\rightarrow\square\psi)$
\end{itemize}
\end{lemma}

The first item of Lemma \ref{124} says that there is exactly one necessary proposition, namely the proposition denoted by $\top$. The second item is the well-known modal law $K$. A further usefull fact is

\begin{lemma}\label{126}
Let $\varphi,\psi\in Fm$. Then: $\vdash\square(\varphi\wedge\psi)\leftrightarrow (\square\varphi\wedge\square\psi)$.
\end{lemma}

\paragraph*{Proof.}
$(\varphi\wedge\psi)\rightarrow\varphi$ and $(\varphi\wedge\psi)\rightarrow\psi$ are (substitution-instances of) intuitionistic theorems. Rule AN together with modal law K and rule MP yield $\vdash\square(\varphi\wedge\psi)\rightarrow \square\varphi$ and $\vdash\square(\varphi\wedge\psi)\rightarrow \square\psi$. By propositional logic, $\vdash\square(\varphi\wedge\psi)\rightarrow (\square\varphi\wedge\square\psi)$. On the other hand, $\varphi\rightarrow (\psi\rightarrow (\varphi\wedge\psi))$ is an intuitionistic theorem. Again, rule AN and principle K yield $\vdash\square\varphi\rightarrow \square(\psi\rightarrow (\varphi\wedge\psi))$. By transitivity of implication and principle K, $\vdash\square\varphi\rightarrow (\square\psi\rightarrow\square (\varphi\wedge\psi))$. Thus, $\vdash (\square\varphi\wedge\square\psi)\rightarrow\square(\varphi\wedge\psi)$. Now, the assertion follows. Q.E.D.\\

By Lemma \ref{126}, strict equivalence modulo L can be written by $\square(\varphi\leftrightarrow\psi)$ instead of $\square(\varphi\rightarrow\psi)\wedge \square(\psi\rightarrow\varphi)$.\\

For propositional formulas $\Phi\cup\{\varphi\}\subseteq Fm_0$, we write $\Phi\vdash_{IPC}\varphi$ if there is a derivation of $\varphi$ from $\Phi$ in intuitionistic propositional logic. Now, we are already able to show the easy part of our Main Theorem:

\begin{lemma}\label{130}
Let $\Phi\cup\{\varphi\}\subseteq Fm_0$. Then $\Phi\vdash_{IPC}\varphi$ implies $\square\Phi\vdash\square\varphi$.
\end{lemma}

\paragraph*{Proof.}
Suppose $\Phi\vdash_{IPC}\varphi$. We show the assertion by induction on a derivation. If $\varphi$ is an intuitionistic axiom, then we apply rule AN to obtain $\square\varphi$. If $\varphi\in\Phi$, then obviously $\square\varphi\in\square\Phi$. Finally, suppose $\varphi$ is obtained in intuitionistic propositional logic by Modus Ponens from $\psi$ and $\psi\rightarrow\varphi$. By induction hypothesis, $\square\Phi\vdash\square\psi$ and $\square\Phi\vdash\square(\psi\rightarrow\varphi)$. Then modal law $K$ and rule MP yield $\square\Phi\vdash\square\varphi$. Q.E.D.

\section{Semantics}

Recall that a Heyting algebra is a bounded lattice $\mathcal{H}=(H,f_\top, f_\bot, f_\vee, f_\wedge)$ with an additional binary operation $f_\rightarrow$ which maps any two elements $m,m'\in H$ to the greatest element $f_\rightarrow(m,m')=m''\in H$ with the property $f_\wedge(m,m'')\le m'$, where $\le$ is the lattice ordering. Given such an operation for implication, an operation $f_\neg$ for complement is defined as follows: $f_\neg(m)=f_\rightarrow(m,f_\bot)$, for any $m\in H$. A Heyting algebra is a Boolean algebra if $f_\rightarrow(m,m')=f_\vee(f_\neg(m),m'))$, for all elements $m,m'$. On the other hand, every Boolean algebra gives rise to a Heyting algebra if one defines an implication operation in that way. A filter of a Heyting algebra $\mathcal{H}$ is a non-empty subset $F\subseteq H$ such that for all $m,m'\in H$ the following hold:\\
(a) $m\in F$ and $m\le m'$ implies $m'\in F$,\\
(b) $m\in F$ and $m'\in F$ implies $f_\wedge(m,m')\in F$,\\
(c) $f_\bot\notin F$.\\
A filter $F$ is prime if for all $m,m'\in H$ the following condition is fulfilled:\\ 
(d) $f_\vee(m,m')\in F$ implies $m\in F$ or $m'\in F$.\\
Finally, an ultrafilter is a filter which is maximal with respect to inclusion. A filter $F$ is an ultrafilter iff it has the following property: $m\notin F$ iff $f_\neg(m)\in F$, for all $m\in H$. Every ultrafilter is prime. If the underlying Heyting algebra is a Boolean algebra, then it also holds the converse: every prime filter is an ultrafilter. The intersection of a non-empty set of filters is a filter. The smallest filter is the set $\{f_\top\}$. It follows from Zorn's Lemma that every filter extends to an ultrafilter.\\

Notice that the smallest filter $\{f_\top\}$ of a Heyting algebra is not necessarily prime. For instance, in any Boolean algebra we have $f_\vee(m,f_\neg(m))=f_\top$, for any element $m$. If this is not the two-element algebra, then we do not necessarily have $m=f_\top$ or $f_\neg(m)=f_\top$ (consider a powerset algebra). On the other hand, the L\"owenheim-Tarski algebra of intuitionistic propositional logic is an example of a Heyting algebra with the disjunction property (in fact, it is well-known that the smallest theory of intuitionistic logic, i.e. the set of intuitionistic theorems, is a prime theory; see, e.g., \cite{dal}).

\begin{definition}\label{132}
Let $\mathcal{H}$ be an Heyting algebra with top element $f_\top$. We say that $\mathcal{H}$ has the disjunction property if the smallest filter $\{f_\top\}$ is prime. 
\end{definition}

In a Heyting algebra with the disjunction property, it holds the following property for all elements $m,m'$: $f_\vee(m,m')=f_\top\Leftrightarrow m=f_\top$ or $m'=f_\top$. There is, up to isomorphism, only one non-trivial Boolean algebra with the disjunction property:

\begin{lemma}\label{136}
Let $\mathcal{B}$ be a Boolean algebra. Then $\mathcal{B}$ has the disjunction property iff $\mathcal{B}$ has at most two elements.
\end{lemma}

\paragraph*{Proof.}
It is clear that any Boolean algebra with no more than two elements has the disjunction property. Suppose a Boolean algebra $\mathcal{B}$ has the disjunction property. Then for all its elements $m$: $f_\vee(m,f_\neg(m))=f_\top$. Thus, $m=f_\top$ or $f_\neg(m)=f_\top$, for all elements $m$. In a Boolean algebra, the latter equation is equivalent with $m=f_\bot$. That is, any element of $\mathcal{B}$ is either the greatest element or it is the smallest element. Q.E.D.

\begin{definition}\label{140}
A model $\mathcal{M}=(M, \mathit{TRUE}, f_\top, f_\bot, f_\rightarrow, f_\vee, f_\wedge, f_\square)$ is a Heyting algebra with an ultrafilter $\mathit{TRUE}\subseteq M$ and an additional operation $f_\square$ such that for all $m,m',m''\in M$ and the induced lattice ordering $\le$ the following truth conditions hold:
\begin{enumerate}
\item $f_\square(m)\le m$
\item $f_\square(f_\rightarrow(m,m'))\le f_\rightarrow(f_\square(f_\rightarrow(m',m'')),f_\square(f_\rightarrow(m,m'')))$
\item $f_\square(f_\vee(m,m'))\le f_\vee(f_\square(m),f_\square(m'))$
\item $f_\square(m)\in\mathit{TRUE}\Leftrightarrow m=f_\top$
\end{enumerate}
The elements of $M$ are called propositions, $\mathit{TRUE}$ is the set of true propositions, and $f_\top$ is the necessary (or the proved) proposition. An assignment on a model $\mathcal{M}$ is a function $\gamma\colon V\rightarrow M$ which extends in the canonical way to a function $\gamma\colon Fm\rightarrow M$, i.e. $\gamma(\bot)=f_\bot$, $\gamma(\square\varphi)=f_\square(\gamma(\varphi))$ and $\gamma(\varphi *\psi)=f_*(\gamma(\varphi),\gamma(\psi))$, where $*\in\{\rightarrow,\vee,\wedge\}$.
\end{definition}

\begin{lemma}\label{150}
Every model has the disjunction property.
\end{lemma}

\paragraph*{Proof.} By conditions (iv) and (iii) of a model and the fact that $\mathit{TRUE}$ is a prime filter, we may argue in the following way: $f_\vee(m,m')=f_\top$ $\Rightarrow$ $f_\square(f_\vee(m,m'))\in\mathit{TRUE}$ $\Rightarrow$ $f_\vee(f_\square(m),f_\square(m'))\in\mathit{TRUE}$ $\Rightarrow$ $f_\square(m)\in\mathit{TRUE}$ or $f_\square(m')\in\mathit{TRUE}$ $\Rightarrow$ $m=f_\top$ or $m'=f_\top$. Q.E.D.

\begin{definition}\label{160}
An interpretation is a tuple $(\mathcal{M},\gamma)$ consisting of a model $\mathcal{M}$ and an assignment $\gamma\colon V\rightarrow M$. The relation of satisfaction (truth) is given by 
\begin{equation*}
(\mathcal{M},\gamma)\vDash\varphi :\Leftrightarrow\gamma(\varphi)\in TRUE.
\end{equation*}
This notion extends in the usual way to sets of formulas. For a set of formulas $\Phi$, we define $Mod(\Phi):=\{(\mathcal{M},\gamma)\mid(\mathcal{M},\gamma)\vDash\Phi\}$. This gives rise to the following relation of logical consequence:  
\begin{equation*}
\begin{split}
\Phi\Vdash\varphi :\Leftrightarrow Mod(\Phi)\subseteq Mod(\{\varphi\}).
\end{split}
\end{equation*}
\end{definition}

Recall that in any Heyting algebra: $m\le m'\Leftrightarrow f_\rightarrow(m',m)=f_\top$, for all elements $m,m'$. We will tacitly make use of this fact. The next result follows readily from the definitions. It says that the defined connective $\equiv$ has actually the intended meaning of an identity connective.

\begin{theorem}\label{180}
$(\mathcal{M},\gamma)\vDash\varphi\equiv\psi\Leftrightarrow\gamma(\varphi)=\gamma(\psi)$.
\end{theorem}

\begin{corollary}(Substitution Principle)\label{200}
For all $\varphi,\psi,\psi'\in Fm$:\\ $\Vdash (\varphi\equiv\psi)\rightarrow(\chi[x:=\varphi]\equiv\chi[x:=\psi])$.
\end{corollary}

\paragraph*{Proof.}
Let $(\mathcal{M},\gamma)\vDash\varphi\equiv\psi$. That is, $\gamma(\varphi)=\gamma(\psi)$. By induction on $\chi$, one shows that $\gamma(\chi[x:=\varphi])=\gamma(\chi[x:=\psi])$. That is, $(\mathcal{M},\gamma)\vDash\chi[x:=\varphi]\equiv\chi[x:=\psi]$. Q.E.D.\\

We say that a formula $\varphi$ is valid if $\varphi$ denotes a true proposition in every Heyting algebra under every assignment. It is well-known that any intuitonistic tautology denotes the top element in any Heyting algebra under any assignment. By Corollary \ref{200}, the same holds true for any substitution-instance of an intuitionistic tautology. Moreover, by the truth conditions of a model, all axioms of the form (ii)--(iv) denote the top element. Thus, all axioms are valid, and rule AN is sound. By Corollary \ref{200}, scheme SP is valid. Since $\mathit{TRUE}$ is an ultrafilter, also all formulas of the form $\varphi\vee\neg\varphi$ are valid (although such a formula does not necessarily denote the top element of a given Heyting algebra). Finally, suppose we are given a Heyting algebra such that $m\in\mathit{TRUE}$ and $f_\rightarrow(m,m')\in\mathit{TRUE}$. Since $\mathit{TRUE}$ is a filter, we have $f_\wedge(m,f_\rightarrow(m,m'))\in\mathit{TRUE}$. If $\le$ is the lattice ordering, then $f_\wedge(m,f_\rightarrow(m,m'))\le m'$, as in any Heyting algebra. Thus, $m'\in\mathit{TRUE}$. This shows that the rule of Modus Ponens is sound. Soundness of the calculus now follows by induction on the length of a derivation.

\begin{theorem}[Soundness]\label{240}
For any set $\Phi\cup\{\varphi\}\subseteq Fm$, $\Phi\vdash\varphi$ implies $\Phi\Vdash\varphi$.
\end{theorem}

\begin{corollary}\label{260}
Logic L is a conservative extension of classical propositional logic. That is, for any $\varphi\in Fm_0$, $\varphi$ is a theorem of CPC iff $\varphi$ is a theorem of L.
\end{corollary}

\paragraph*{Proof.}
By definition of L as a deductive system, it contains all classical propositional tautologies as theorems. Thus, L extends CPC. Now let $\varphi$ be a propositional formula with $\vdash\varphi$. By soundness, $\varphi$ is valid in L. In particular, $\varphi$ denotes the unique true proposition (the top element) of the two-element Boolean algebra, under any assignment. This means that $\varphi$ evaluates to \textit{true} under any boolean, i.e. bivalent, truth-value assignment. Thus, $\varphi$ is valid in CPC. Q.E.D.

\section{Completeness}

The notions of a consistent, maximally consistent, inconsistent set of formulas w.r.t. the deductive system L are defined in the usual way. A theory is a set of formulas which is consistent and deductively closed. For a maximal theory $\Phi$, i.e. a maximally consistent set of formulas $\Phi$, we define 
\begin{equation*}
\varphi\approx_\Phi\psi :\Leftrightarrow\Phi\vdash\varphi\equiv\psi.
\end{equation*}

Since L extends CPC and is itself a classical logic, a maximal theory $\Phi$ has the well-known properties of a maximally consistent set such as $\varphi\in\Phi$ iff $\Phi\vdash\varphi$, $\Phi\vdash\varphi$ or $\Phi\vdash\neg\varphi$, for any $\varphi$, etc. In particular, $\Phi$ is a prime theory.

\begin{lemma}\label{320}
Let $\Phi$ be a maximal theory. Then $\approx_\Phi$ is an equivalence relation on $Fm$ with the following properties:
\begin{itemize}
\item If $\varphi_1\approx_\Phi\psi_1$ and $\varphi_2\approx_\Phi\psi_2$, then $\square\varphi_1\approx_\Phi\square\psi_1$ and $(\varphi_1 * \varphi_2)\approx_\Phi(\psi_1 * \psi_2)$, where $*\in\{\vee,\wedge,\rightarrow\}$.
\item If $\varphi\approx_\Phi\psi$, then $\varphi\in\Phi\Leftrightarrow\psi\in\Phi$.
\item If $\varphi\approx_\Phi\psi$, then $\square\varphi\in\Phi\Leftrightarrow\square\psi\in\Phi$.
\end{itemize}
\end{lemma}

\paragraph*{Proof.}
It follows from axioms of (intuitionistic) propositional logic and rule AN that $\approx_\Phi$ is reflexive and symmetric. Transitivity follows from applications of axiom (iii). Thus, $\approx_\Phi$ is an equivalence relation on $Fm$. Now suppose $\varphi_1\approx_\Phi\psi_1$ and $\varphi_2\approx_\Phi\psi_2$. Let $x\neq y$ be variables such that $x$ does not occur in $\psi_2$ and $y$ does not occur in $\varphi_1$. Then by SP and rule MP: $(\varphi_1 * \varphi_2)=(\varphi_1 * y)[y:=\varphi_2]\approx_\Phi(\varphi_1 * y)[y:=\psi_2]=(\varphi_1 * \psi_2)=(x * \psi_2)[x:=\varphi_1]\approx_\Phi(x * \psi_2)[x:=\psi_1]=(\psi_1 * \psi_2)$. By transitivity of $\approx_\Phi$, we get $(\varphi_1 * \varphi_2)\approx_\Phi(\psi_1 * \psi_2)$. Similarly, one shows $\square\varphi_1\approx_\Phi\square\psi_1$. The second item of the Lemma follows from axiom (ii) and MP. The third item follows from the previous items of the Lemma. Q.E.D.

\begin{lemma}\label{330}
Every consistent set is satisfiable.
\end{lemma}

\paragraph*{Proof.}
By Zorn's Lemma (or even by weaker assumptions), a consistent set $\Psi$ extends to a maximal theory $\Phi$. For $\varphi\in Fm$, let $\overline{\varphi}$ be the equivalence class of $\varphi$ modulo $\approx_\Phi$. We define a model $\mathcal{M}$ with the following ingredients:
\begin{itemize}
\item $M:=\{\overline{\varphi}\mid\varphi\in Fm\}$ 
\item $\mathit{TRUE}:=\{\overline{\varphi}\mid\varphi\in \Phi\}$
\item functions $f_\top, f_\bot$, $f_\square$, $f_*$, where $*\in\{\vee,\wedge,\rightarrow\}$, defined by $f_\top:=\overline{\top}$, $f_\bot:=\overline{\bot}$, $f_\square(\overline{\varphi}):=\overline{\square\varphi}$, $f_*(\overline{\varphi},\overline{\psi}):=\overline{\varphi * \psi}$, respectively.
\end{itemize}

By Lemma \ref{320}, all these ingredients are well-defined. We show that $\mathcal{M}$ has the properties of a model established in Definition \ref{140}. A Heyting algebra can be characterized as a bounded lattice with an operation $f_\rightarrow$ such that certain equations are satisfied such as $f_\rightarrow(m,m)=f_\top$, etc. All these equations are interpretations of intuitionistic theorems which are of the form of biconditionals such as $(\varphi\rightarrow\varphi)\leftrightarrow\top$, etc. Since $\Phi$ is a theory, it contains all intuitionistic theorems, particularly those of the form $\varphi_1\leftrightarrow\varphi_2$. By rule AN, $\Phi$ then contains $\square(\varphi_1\leftrightarrow\varphi_2)$. Hence, $\overline{\varphi_1}=\overline{\varphi_2}$. This shows that $\mathcal{M}$ satisfies all equational axioms which characterize the class of Heyting algebras. Recall that the underlying lattice ordering $\le$ can be recovered by the equivalence $\overline{\varphi}\le\overline{\psi}$ $\Leftrightarrow$ $\overline{\varphi\rightarrow\psi}=f_\top$. Now one easily checks that $\mathit{TRUE}$ is an ultrafilter on $\mathcal{M}$. By Lemma \ref{124}, $\square\varphi\in\Phi\Leftrightarrow\varphi\equiv\top\in\Phi$. Thus, $f_\square(\overline{\varphi})\in\mathit{TRUE}\Leftrightarrow\overline{\varphi}=\overline{f_\top}$. That is, truth condition (iv) of a model (see Definition \ref{140}) is satisfied. $\Phi$ contains, in particular, the axioms (ii)--(iv). By applying rule AN one shows that also the truth conditions (i)--(iii) of a model are satisfied. Thus, $\mathcal{M}$ is a model in the sense of Definition \ref{140}. We consider the assignment $\gamma\colon V\rightarrow M$ defined by $x\mapsto\overline{x}$. Then by induction on formulas one easily shows that $\gamma(\varphi)=\overline{\varphi}$, for any formula $\varphi$. Thus:
$\varphi\in\Phi\Leftrightarrow\overline{\varphi}\in \mathit{TRUE}\Leftrightarrow\gamma(\varphi)\in \mathit{TRUE}\Leftrightarrow (\mathcal{M},\gamma)\vDash\varphi$. Q.E.D.\\

In the same way as in CPC, $\Phi\nvdash\varphi$ implies that the set $\Phi\cup\{\neg\varphi\}$ is consistent. The Completeness Theorem then follows from Lemma \ref{330}.

\begin{corollary}[Completeness Theorem]\label{350}
Let $\Phi\cup\{\varphi\}\subseteq Fm$. Then $\Phi\Vdash\varphi$ implies $\Phi\vdash\varphi$.
\end{corollary}

\section{Main Theorem and Conclusions}

G\"odel's result \cite{goe} which says that $\vdash_{IPC}\varphi$ implies $\vdash_{S4}\varphi'$, where the map $\varphi\mapsto\varphi'$ is G\"odel's translation, can be shown by induction on derivations in IPC. G\"odel's conjecture that also the converse is true is harder to prove and was first established years later by McKinsey and Tarski \cite{mcktar}. Fortunately, things are somewhat less complicated in the case of our embedding of IPC into logic L which is managed by the simpler map $\varphi\mapsto\square\varphi$. The proof that $\Phi\vdash_{IPC}\varphi$ implies $\Phi\vdash_L\square\varphi$ relies on a straightforward induction on derivations in IPC, see Lemma \ref{130}. In this section, we will show that for propositional $\Phi,\varphi$ the converse holds, too.

\begin{theorem}\label{400}
Let $\Phi\cup\{\chi\}\subseteq Fm_0$. Then $\square\Phi\vdash_L\square\chi\Leftrightarrow\Phi\vdash_{IPC}\chi$.
\end{theorem}

\paragraph*{Proof.}
The direction from right to left is Lemma \ref{130}. Suppose $\Phi\nvdash_{IPC}\chi$. By a standard construction (see, e.g., \cite{dal}), we may extend $\Phi$ to a prime theory $\Phi_p\subseteq Fm_0$ of IPC such that $\Phi_p\nvdash_{IPC}\chi$. By a standard application of Zorn's Lemma, $\Phi_p$ is contained in a maximal theory $\Phi_{max}\subseteq Fm_0$ of IPC. Then $\Phi_{max}$ is also a maximal theory of CPC. In fact, a Kripke model of $\Phi_{max}$ is a singleton, i.e. a classical truth-value assignment. Note that possibly $\chi\in\Phi_{max}$. We now construct a model of $\Phi_{max}$ that identifies precisely those formulas $\varphi,\psi\in Fm_0$ which are intuitionistically equivalent modulo prime theory $\Phi_p\subseteq \Phi_{max}$. For this purpose, we define the relation $\approx$ on $Fm_0$ by 
\begin{equation*}
\varphi\approx\psi :\Leftrightarrow\Phi_p\vdash_{IPC}\varphi\leftrightarrow\psi. 
\end{equation*}
Notice that $\approx$ is defined on $Fm_0\subseteq Fm$ and not on the whole set $Fm$. By IPC, $\approx$ is a congruence on $Fm_0$, i.e. $\approx$ is an equivalence relation and $\varphi_1\approx\psi_1$, $\varphi_2\approx\psi_2$ imply $(\varphi_1 * \varphi_2)\approx(\psi_1 *\psi_2)$ for $*\in\{\vee,\wedge,\rightarrow\}$. Let $\overline{\varphi}$ denote the congruence class of $\varphi\in Fm_0$ modulo $\approx$. We define
\begin{equation*}
\begin{split}
&M:=\{\overline{\varphi}\mid \varphi\in Fm_0\}\\
&\mathit{TRUE}:=\{\overline{\varphi}\mid\varphi\in\Phi_{max}\}\\
&f_\bot:=\overline{\bot}\\
&f_\top:=\overline{\bot\rightarrow\bot}\\
&f_*(\overline{\varphi},\overline{\psi}):=\overline{\varphi * \psi}\text{ for }*\in\{\vee,\wedge,\rightarrow\}\\
&f_\square(\overline{\varphi}):=
\begin{cases}
&f_\top,\text{ if }\overline{\varphi}=f_\top\\
&f_\bot,\text{ else. }
\end{cases}
\end{split}
\end{equation*}
Since $\approx$ is a congruence on $Fm_0$, these operations are well-defined. It is clear that $\varphi,\psi\in\Phi_p$ implies $\Phi_p\vdash_{IPC}\varphi\leftrightarrow\psi$. It follows that $f_\top=\Phi_p$. Recall that Heyting algebras can be axiomatized by equations that correspond to theorems of IPC which are given in the form $\varphi\leftrightarrow\psi$. Then by construction, the above defined operations form a Heyting algebra on $M$ with ultrafilter $\mathit{TRUE}$ and an operation $f_\square$. It remains to show that the truth conditions (i)--(iv) of Definition \ref{140} are satisfied. Condition (iv) follows readily from the definition of $f_\square$. Let us look at condition (i), $f_\square(\overline{\varphi})\le\overline{\varphi}$, for any $\overline{\varphi}\in M$, where $\le$ is the underlying lattice ordering. By definition, there are only two possibilities: $f_\square(\overline{\varphi})=f_\top$ or $f_\square(\overline{\varphi})=f_\bot$. In the latter case, there is nothing to show, since $f_\bot$ is the bottom element of the Heyting lattice. We assume the former case. Then, by definition of $f_\square$, we get $\overline{\varphi}=f_\top$. Hence, condition (i) is satisfied. Now, we consider condition (ii), $f_\square(f_\rightarrow(\overline{\varphi},\overline{\psi}))\le f_\rightarrow(f_\square(f_\rightarrow(\overline{\psi},\overline{\chi})),f_\square(f_\rightarrow(\overline{\varphi},\overline{\chi})))$. Again, if $f_\square(f_\rightarrow(\overline{\varphi},\overline{\psi}))=f_\bot$, there is nothing to show. So we may assume $f_\square(f_\rightarrow(\overline{\varphi},\overline{\psi}))=f_\top$, i.e. $\overline{\varphi}\le \overline{\psi}$. Moreover, we may assume $f_\square(f_\rightarrow(\overline{\psi},\overline{\chi}))=f_\top$, since otherwise we are ready. This implies $\psi\le \chi$. Transitivity of $\le$ yields $\varphi\le \chi$, i.e. $f_\square(f_\rightarrow(\overline{\varphi},\overline{\chi})))=f_\top$. It follows that $f_\rightarrow(f_\square(f_\rightarrow(\overline{\psi},\overline{\chi})),f_\square(f_\rightarrow(\overline{\varphi},\overline{\chi})))=f_\top$. Hence, condition (ii) is verified. Finally, let us check condition (iii), $f_\square(f_\vee(\overline{\varphi},\overline{\psi}))\le f_\vee(f_\square(\overline{\varphi}),f_\square(\overline{\psi}))$. Suppose $f_\square(f_\vee(\overline{\varphi},\overline{\psi}))=f_\top$. Then $f_\vee(\overline{\varphi},\overline{\psi})=f_\top=\Phi_p$. Thus, $\overline{\varphi\vee\psi}=\Phi_p$ and therefore $\varphi\vee\psi\in\Phi_p$. Since $\Phi_p$ is a prime theory, $\varphi\in \Phi_p$ or $\psi\in\Phi_p$. This means that $\overline{\varphi}=f_\top$ or $\overline{\psi}=f_\top$. That is, $f_\square(\overline{\varphi})=f_\top$ or $f_\square(\overline{\psi})=f_\top$. In any case, $f_\vee(f_\square(\overline{\varphi}),f_\square(\overline{\psi}))=f_\top$, and condition (iii) is satisfied. We have shown that $\mathcal{M}$ is a model in the sense of Definition \ref{140}.  

Now, we consider the assignment $\varepsilon\colon V\rightarrow M$ defined by $x\mapsto\overline{x}$. By induction, $\varepsilon(\varphi)=\overline{\varphi}$, for any formula $\varphi\in Fm_0$. By construction, $\Phi\subseteq\Phi_p$ and $\chi\notin\Phi_p$. Furthermore, for every $\varphi\in Fm_0$:
\begin{equation*}
(\mathcal{M},\varepsilon)\vDash\square\varphi\Leftrightarrow\varepsilon(\square\varphi)=f_\square(\overline{\varphi})\in \mathit{TRUE}\Leftrightarrow\overline{\varphi}=f_\top=\Phi_p\Leftrightarrow\varphi\in\Phi_p.
\end{equation*}
Hence, $(\mathcal{M},\varepsilon)\vDash\square\Phi$ and $(\mathcal{M},\varepsilon)\nvDash\square\chi$. Thus, $\square\Phi\nVdash\square\chi$. Soundness yields $\square\Phi\nvdash\square\chi$. Q.E.D.\\

In Kripke semantics of IPC, truth means satisfaction at a given possible world. The set of all formulas which are true at a given world form a prime theory of IPC. Thus, intuitionistic truth of a formula $\varphi$ means that $\varphi$ belongs to a given prime theory of IPC. The next two results show that the modal operator, restricted to propositional formulas, is a predicate for intuitionistic truth. In this sense, the  modal operator of L can be viewed as a provability predicate. 

\begin{corollary}\label{406}
For any prime theory $\Phi_p\subseteq Fm_0$ of IPC there is an interpretation $(\mathcal{M},\varepsilon)$ such that for all $\varphi\in Fm_0$:
\begin{equation*}
(\mathcal{M},\varepsilon)\vDash\square\varphi\Leftrightarrow\varphi\in \Phi_p.
\end{equation*}
\end{corollary}

\paragraph*{Proof.}
This follows immediately from the construction given in the proof of Theorem \ref{400}. Q.E.D.

\begin{proposition}\label{408}
For every interpretation $(\mathcal{M},\gamma)$ there is a unique prime theory $\Phi_p\subseteq Fm_0$ of IPC such that for all $\varphi\in Fm_0$:
\begin{equation*}
 (\mathcal{M},\gamma)\vDash\square\varphi\Leftrightarrow\varphi\in\Phi_p.
\end{equation*}
That is, the modal operator $\square$, restricted to propositional formulas, can be seen as a predicate for intuitionistic truth.
\end{proposition}

\paragraph*{Proof.}
Let $\Phi_p:=\{\varphi\in Fm_0\mid \gamma(\varphi)=f_\top\}$. Then by truth conditions of a model: $(\mathcal{M},\gamma)\vDash\square\varphi$ iff $\gamma(\square\varphi)\in\mathit{TRUE}$ iff $f_\square(\gamma(\varphi))\in\mathit{TRUE}$ iff $\gamma(\varphi)=f_\top$ iff $\varphi\in\Phi_p$, for any $\varphi\in Fm_0$. We show by induction on a derivation that $\Phi_p\vdash_{IPC}\psi$ implies $\psi\in\Phi_p$, for any $\psi\in Fm_0$. If $\psi$ is an intuitionistic axiom, then it denotes the top element $f_\top$ of the Heyting algebra, i.e. $\gamma(\psi)=f_\top$ and $\psi\in\Phi_p$. Finally, suppose $\psi$ is obtained by Modus Ponens from formulas $\varphi$ and $\varphi\rightarrow\psi$. By induction hypothesis, $\gamma(\varphi)= f_\top$ and $\gamma(\varphi\rightarrow\psi)= f_\top$. The latter means $\gamma(\varphi)\le\gamma(\psi)$, where $\le$ is the lattice ordering. Thus, $\gamma(\psi)=f_\top$ and $\psi\in\Phi_p$. We have shown that $\Phi_p$ is deductively closed in IPC. Then follows that $\Phi_p$ is also consistent in IPC. Thus, $\Phi_p$ is an intuitionistic theory. Suppose $\varphi\vee\psi\in \Phi_p$. Then $\gamma(\varphi\vee\psi)=f_\vee(\gamma(\varphi),\gamma(\psi))=f_\top$. Since every model has the disjunction property, $\gamma(\varphi)=f_\top$ or $\gamma(\psi)=f_\top$. That is, $\varphi\in\Phi_p$ or $\psi\in\Phi_p$, and $\Phi_p$ is a prime theory. It is clear that $\Phi_p$ is unique with the asserted property. Q.E.D.\\

It is known that the set of intuitionistic theorems is a prime theory. This fact is called the \textit{disjunction property} of intuitionistic logic (see, e.g., \cite{dal}). We conclude:

\begin{corollary}\label{410}
There is an interpretation $(\mathcal{M}_{IPC},\varepsilon)$ such that for all propositional $\varphi\in Fm_0$:
\begin{equation*}
(\mathcal{M}_{IPC},\varepsilon)\vDash\square\varphi\Leftrightarrow\text{ }\vdash_{IPC}\varphi.
\end{equation*}
\end{corollary}

By Lemma \ref{126}, $\vdash\varphi\equiv\psi$ iff $\vdash\square(\varphi\leftrightarrow\psi)$. Then Corollary \ref{410} (or, more directly, the model construction given in the proof of Theorem \ref{400}) implies the following.

\begin{corollary}\label{420}
There is an interpretation $(\mathcal{M}_{IPC},\varepsilon)$ such that for all $\varphi,\psi\in Fm_0$,
\begin{equation*}
(\mathcal{M}_{IPC},\varepsilon)\vDash\varphi\equiv\psi\Leftrightarrow\text{}\vdash_{IPC}\varphi\leftrightarrow\psi.
\end{equation*}
\end{corollary}

That is, model $(\mathcal{M}_{IPC},\varepsilon)$ identifies exactly those propositional formulas which are intuitionistically equivalent. In general, however, a model satisfies more equations. In the two-element Boolean algebra, for example, the equations $\varphi\equiv\neg\neg\varphi$ and $\varphi\equiv\square\varphi$ are true. A result stronger than Corollary \ref{420} would be the existence of a model $(\mathcal{M},\gamma)$ with the following property. For all $\varphi,\psi\in Fm$:
\begin{equation*}
(\mathcal{M},\gamma)\vDash\varphi\equiv\psi\Leftrightarrow\text{}\vdash_L\varphi\equiv\psi.
\end{equation*}
We call such an interpretation a \textit{canonical model}. A canonical model satisfies only those equations which are satisfied in all models. We have no proof for the existence of such a model for logic L and leave this question as an open problem. The models of CPC can be regarded as two-element Boolean algebras. Thus, propositional identity $\varphi\equiv\psi$ collapses with material equivalence $\varphi\leftrightarrow\psi$ in CPC. Then it is clear that a canonical model cannot exist. In IPC, propositional identity collapses with intuitionistic equivalence. In fact, if we regard a proposition denoted by $\varphi$ as the set of possible worlds where $\varphi$ is true and which are accessible from the current world $w$ (see, e.g., \cite{lewsl}), then $\varphi\leftrightarrow\psi$ is true at $w$ iff at each accessible world,  both formulas are true or both formulas are not true. This is the same as to say that (at the actual world $w$) $\varphi$ and $\psi$ denote the same proposition: $\varphi\equiv\psi$. A canonical model for IPC is the root of the Kripke frame consisting of all prime theories as possible worlds and settheoretic inclusion as accessibility relation. The root $w_0$, i.e. the smallest prime theory, is the set of intuitionistic theorems. Then $\varphi\leftrightarrow\psi$ is true at $w_0$ iff $\vdash_{IPC}\varphi\leftrightarrow\psi$.

It is not hard to construct a canonical model for Suszko's basic non-Fregean logic SCI \cite{blosus} such that $\varphi\equiv\psi$ is satisfied iff $\varphi=\psi$ (see \cite{lewnd}). Consequently, $\vdash_{SCI}\varphi\equiv\psi$ iff $\varphi=\psi$. In \cite{lewigpl}, a canonical model $\mathcal{M}$ for a non-Fregean logic with propositional quantifiers is constructed such that for any two sentences $\varphi$ and $\psi$ it holds that $\mathcal{M}\vDash\varphi\equiv\psi$ iff $\varphi$ and $\psi$ differ at most on bound variables.

\end{document}